%% file: main.tex
\documentclass[submission,copyright,creativecommons]{eptcs}
\providecommand{\event}{PLACES 2019} 
\usepackage{breakurl}             
\usepackage{underscore}           
\usepackage{xcolor}
\usepackage{graphicx}
\usepackage{etoolbox}
\usepackage{amssymb, amsmath}
\usepackage{bbold}
\usepackage{stmaryrd}
\usepackage{xspace}
\usepackage{upgreek}

\input{macros}

\definecolor{mymagenta}{rgb}{0.5,0,0.5}
\definecolor{myred}{rgb}{0.5,0,0}
\definecolor{mygreen}{rgb}{0,0.4,0}
\definecolor{myblue}{rgb}{0,0,0.6}
\definecolor{highlight}{gray}{0.95}

\usepackage{listings}
\lstdefinestyle{INLINE}{
  basicstyle=\tt,
}
\lstdefinestyle{DISPLAY}{
  escapechar=\%,
  numbers=left,
  basicstyle=\tt\small,
  numberstyle=\tt\tiny\color{gray},
  numbersep=0.5em,
  xleftmargin=1em,
  frame=lines,
  backgroundcolor=\color{highlight},
}
\lstdefinestyle{FLOAT}{
  float,
  captionpos=b,
  abovecaptionskip=1em,
  numberbychapter=true,
}
\lstdefinelanguage{MYJAVA}{
  language=Java,
  keywordstyle=\color{myblue},
  commentstyle=\color{mygreen},
  emph={@Reaction,@Protocol,@Operation,@State},
  emphstyle=\color{mymagenta},
  showstringspaces=false,
  columns=fixed,
  mathescape=true,
  basewidth=0.5em,
  literate={·}{{$\tand$}}1
}
\newcommand{\JI}{\lstinline[language=MYJAVA,style=INLINE]}
\lstnewenvironment{JD}[1][]{\lstset{language=MYJAVA,style=DISPLAY,#1}}{}
\lstnewenvironment{JF}[1][]{\lstset{language=MYJAVA,style=DISPLAY,style=FLOAT,#1}}{}

\title{Concurrent Typestate-Oriented Programming in Java}
\author{Rosita Gerbo \qquad Luca Padovani
  \institute{Dipartimento di Informatica, Universit\`a di Torino, Italy}
}
\def\titlerunning{Concurrent Typestate-Oriented Programming in Java}
\def\authorrunning{Rosita Gerbo and Luca Padovani}
\begin{document}
\maketitle

\begin{abstract}
  We describe a generative approach that enables concurrent
  typestate-oriented programming in Java and other mainstream
  languages. The approach allows programmers to implement objects
  exposing a state-sensitive interface using a high-level
  synchronization abstraction that synchronizes methods with the
  states of the receiver object in which those methods have an
  effect. An external tool takes care of generating all the
  boilerplate code that implements the synchronization
  logic. Behavioral types are used to specify object protocols. The
  tool integrates protocol conformance verification with the
  synchronization logic so that protocol violations are promptly
  detected at runtime.
\end{abstract}

\input{introduction}

\input{example}

\input{automaton}

\input{generator}

\input{conclusion}

\paragraph{Acknowledgments.} We are grateful for the valuable and
detailed feedback received from the anonymous reviewers. This work
has been partially supported by MSCA-RISE-2017 778233 BEHAPI.

\bibliographystyle{eptcs}
\bibliography{main}
\end{document}

\appendix

\section{Appendix}

Figure~\ref{fig:untyped.future} shows the matching automaton
corresponding to the join pattern~\eqref{eq:future}, but where the
type of the object is
\[
  \tstar(\TagEMPTY \tor \TagFULL \tor \TagGet \tor \TagPut)
\]
instead of $\FutureType$. This type allows for \emph{any}
combination of messages to be sent to the object, which makes the
join pattern effectively untyped.  The automaton is not meant to be
``comprehensible'' and is shown here only to witness the complexity
of the automata obtained through Le Fessant and Maranget's
construction, even for relatively simple object definitions.

\begin{figure}
  \begin{center}
    \includegraphics[angle=90,height=\textheight]{UntypedFuture.pdf}
  \end{center}
  \caption{\label{fig:untyped.future} Matching automaton for the
    join patterns of the future variable (untyped version).}
\end{figure}

\end{document}

%% file: macros.tex
\newcommand{\TagReply}{\Tag[reply]}

\newcommand{\TagEMPTY}{\Tag[EMPTY]}
\newcommand{\TagFULL}{\Tag[FULL]}
\newcommand{\TagPut}{\Tag[put]}
\newcommand{\TagGet}{\Tag[get]}

\newif\ifbox
\boxfalse

\newif\ifproofs
\proofstrue
\proofsfalse

\newif\ifcomments
\commentstrue

\newif\iflong
\longtrue
\longfalse

\definecolor{mymagenta}{rgb}{0.5,0,0.5}
\definecolor{myred}{rgb}{0.5,0,0}
\definecolor{mygreen}{rgb}{0,0.4,0}
\definecolor{myblue}{rgb}{0,0,0.6}
\definecolor{highlight}{gray}{0.9}

\newcommand{\REVISED}[1]{#1}

\newcommand{\jc}{{Join Calculus}\xspace}
\newcommand{\ojc}{{Objective} \jc}
\newcommand{\comega}{$\mathsf{C}\upomega$\xspace}

\newcommand{\jocaml}{JoCaml\xspace}

\newcommand{\eg}{\textit{e.g.}\xspace}
\newcommand{\cf}{\textit{cf.}\xspace}
\newcommand{\etal}{\textit{et al.}\xspace}

\newcommand{\natset}{\mathbb{N}}

\newcommand{\counter}{\alpha}

\newcommand{\mkkeyword}[1]{\mathtt{\color{myblue}#1}}

\newcommand{\set}[2][]{%
  \{\,#2\,\}%
  \ifblank{#1}{}{_{#1}}%
}

\newcommand{\ttparens}[1]{\ttlparen#1\ttrparen}
\newcommand{\ttbracks}[1]{\ttlbrack#1\ttrbrack}

\newcommand{\ttbar}{\texttt{\upshape|}}

\newcommand{\ttlparen}{\texttt{\upshape(}}
\newcommand{\ttrparen}{\texttt{\upshape)}}

\newcommand{\ttlbrack}{\texttt{\upshape[}}
\newcommand{\ttrbrack}{\texttt{\upshape]}}

\newcommand{\ttemark}{\texttt{\upshape!}}
\newcommand{\ttampersand}{\texttt{\upshape\&}}

\newenvironment{lines}[1][b]{
  \begin{array}[#1]{@{}l@{}}
}{
  \end{array}
}

\newcommand{\Process}{\ProcessP}
\newcommand{\ProcessP}{P}

\newcommand{\Pattern}{\PatternJ}
\newcommand{\PatternJ}{J}

\newcommand{\Tag}[1][m]{\mathtt{#1}}

\newcommand{\parop}{\mathrel\ttampersand}
\newcommand{\invoke}[2]{#1\ttemark#2}
\newcommand{\sendx}[3]{\invoke{#1}{\msg{#2}{#3}}}

\newcommand{\objdef}[2]{#1\ttbracks{#2}}
\newcommand{\objx}[2]{\mkkeyword{object}\objdef{#1}{#2}}
\newcommand{\obj}[2]{\objx{~#1}{#2}}

\newcommand{\reaction}[2]{#1 \triangleright #2}

\newcommand{\orc}{\mathrel\ttbar}

\newcommand{\msg}[2]{
  #1%
  \ifblank{#2}{}{\ttparens{#2}}%
}

\newcommand{\Signature}{\Sigma}

\newcommand{\tzero}{\mathbb{0}}
\newcommand{\tone}{\mathbb{1}}
\newcommand{\tstar}[1]{{*}#1}
\newcommand{\tor}{+}
\newcommand{\tand}{\cdot}

\newcommand{\E}{\mathsf{E}}
\newcommand{\F}{\mathsf{F}}

\newcommand{\sem}[1]{\llbracket#1\rrbracket}
\newcommand{\der}[2]{#1[#2]}

\newcommand{\eqdef}{\stackrel{\text{\tiny\smash{def}}}=}



%% file: introduction.tex
\section{Introduction}
\label{sec:introduction}

Beckman \etal~\cite{BeckmanKimAldrich11} report that objects with a
state-dependent interface are common in Java applications.  Typical
examples of such objects are \emph{iterators} (which can be advanced
only if they are not finished), \emph{files} (which can be read or
written only when they are open), and \emph{locks} (which can be
released only if they have been previously acquired).
Implementing and using objects with a state-dependent interface is
difficult and error prone, to the point that researchers have
investigated specific methodologies -- such as
\emph{typestate-oriented programming} (TSOP for
short)~\cite{AldrichEtAl09, GarciaEtAl14} -- to help software
development.
TSOP is based on a set of programming abstractions supporting the
design of objects with state-dependent interfaces and a behavioral
type system ensuring that methods invoked on an object belong to the
interface corresponding to that object's state.
Crafa and Padovani~\cite{CrafaPadovani17} have extended TSOP to a
concurrent setting, showing that the Objective Join
Calculus~\cite{FournetLaneveMarangetRemy03} is a natural formal
model for (concurrent) TSOP because of its built-in support for
\emph{join patterns}. In fact, join patterns make it possible not
only to \emph{explicitly associate methods with states} (which is
one of the disinguishing features of TSOP as conceived by Aldrich
\etal~\cite{AldrichEtAl09}) but also to \emph{synchronize methods
  and states}: the process invoking a method that is not available
in an object's current state is suspended until the object moves
into a state for which that method is available again.

While the feasibility of Crafa and Padovani's approach to concurrent
TSOP is partially witnessed by a proof-of-concept type checker for
the Objective Join Calculus~\cite{Padovani18,CobaltBlue}, applying
it to mainstream programming languages presents two substantial
problems. (P1) The few languages that feature built-in join patterns
-- notably \jocaml~\cite{FournetLeFessantMarangetSchmitt03},
\comega~\cite{BentonCardelliFournet04,Comega}, Join
Java~\cite{ItzsteinJasiunas03} and
JErlang~\cite{PlociniczakEisenbach2010} -- are mostly experimental
languages serving specialized communities and/or have not been
maintained for a long time.  Library implementations of join
patterns
\cite{Russo07,Russo08,HallerVanCutsem08,SulzmannLamVanWeert08,TuronRusso11}
have similar issues.
(P2) Retrofitting Crafa and Padovani's type system into an existing
language is conceptually and technically challenging. The ongoing
efforts on the implementation of Linear
Haskell~\cite{BernardyEtAl18} show that this is the case even when
retrofitting a streamlined substructural type system in the
controlled setting of a pure functional language.

\REVISED{The main contribution of this work is a practical approach
  that makes} concurrent TSOP immediately applicable to Java and, in
fact, to virtually every programming language. The approach is
necessarily based on compromises.
We address problem (P1) using \emph{code generation}: the programmer
writing a Java class using our approach specifies the join patterns
that synchronize states and methods by means of \emph{standard Java
  annotations}~\cite{GoslingEtAl14}; an external tool generates the
boilerplate code that implements all the synchronization logic. As a
result, we avoid any dependency on join patterns in the programming
language or in external libraries.
We address problem (P2) trading \emph{compile-time} with
\emph{runtime} protocol conformance verification. The programmer
provides a protocol specification for the class, again in the form
of a Java annotation, and the same external tool generates the code
that detects all protocol violations at runtime if (and as soon as)
they actually occur. This way, we bypass the non-trivial problem of
reconciling fundamentally different type systems, at the cost of
delayed detection of programming errors.
\REVISED{From a technical standpoint, we also contribute a
  refinement of Le Fessant and Maranget's~\cite{LeFessantMaranget98}
  compilation scheme for join patterns that uses behavioral types to
  identify object protocol violations and to prune the state space
  of the automaton -- called \emph{matching automaton} -- that
  performs join pattern matching.}

The rest of the paper is organized as follows.  We start recalling
the key features of concurrent TSOP with join
patterns~\cite{CrafaPadovani17} through a simple example of
concurrent object (Section~\ref{sec:example}). Then, we describe the
construction of the matching automaton
(Section~\ref{sec:automaton}), which is instrumental to the
subsequent code generation phase (Section~\ref{sec:generator}).  We
conclude in Section~\ref{sec:conclusion}.

The tool has been implemented and used to generate all the code and
automata shown in the paper. Its source code is publicly
available~\cite{EasyJoin}.


%% file: example.tex
\section{Example}
\label{sec:example}

\newcommand{\x}{x}
\newcommand{\y}{y}
\newcommand{\Lock}{\mathit{lock}}
\newcommand{\Future}{\mathit{future}}
\newcommand{\LockType}{\E_{\Lock}}
\newcommand{\FutureType}{\E_{\Future}}
\newcommand{\User}{\mathit{user}}

We illustrate the basics of concurrent TSOP in the Objective Join
Calculus~\cite{FournetLaneveMarangetRemy03,CrafaPadovani17} by
modeling a \REVISED{\emph{promise} or \emph{completable future
    variable} (the terminology varies depending on the language).}
We can think of a completable future variable as a one-place buffer
that must be written (or completed) once \REVISED{(this is the view
  of the variable that is sometimes called
  \emph{promise}~\cite{ScalaPromises})} and that can be read any
number of times, concurrently. \REVISED{We find this specific
  example appropriate because it is a relatively simple concurrent
  object that exposes a state-sensitive interface and whose protocol
  uses all the connectives of the type language we are about to
  discuss.}
Below is the definition of a $\Future$ object of type $\FutureType$:

\begin{equation}
  \label{eq:future}
  \begin{lines}[c]
    \obj{\Future : \FutureType
      \begin{array}[t]{@{}l@{~}l@{~}l@{~}l@{~}l@{~}l@{}}
      }{
        &
        \reaction{
          \TagEMPTY & \parop & \msg\TagPut\x &
        }{&
          \sendx\Future\TagFULL\x
        }
        \\
        \orc &
        \reaction{
          \msg\TagFULL\x & \parop & \msg\TagGet\User &
        }{&
          \sendx\Future\TagFULL\x \parop \sendx\User\TagReply\x
        }
        ~
      }
    \end{array}
    \\
    \sendx\Future\TagEMPTY{}
  \end{lines}
\end{equation}

Messages sent to the $\Future$ object are stored into its
\emph{mailbox}. The object understands four kinds of messages, each
identified by a \emph{tag}: $\TagEMPTY$ and $\TagFULL$ model the
state of the object while $\TagPut$ and $\TagGet$ its operations.
Thus, an $\TagEMPTY$ message in the $\Future$'s mailbox (denoted by
a term $\sendx\Future\TagEMPTY{}$) means that $\Future$ has not been
completed yet, whereas a $\msg\TagFULL\x$ message in the $\Future$'s
mailbox means that $\Future$ has been completed with value $\x$.
The behavior of the object is given by the two reactions
$\reaction\Pattern\Process$ within brackets. When (some of) the
messages in the object's mailbox match the join pattern $\Pattern$
of a reaction, the matched messages are atomically consumed and
those on the right hand side of the reaction are produced.
Above, the first reaction specifies that a $\Future$ in state
$\TagEMPTY$ accepts a $\TagPut$ operation carrying an $\x$ argument
and changes its state to $\msg\TagFULL\x$.
The second reaction specifies that a $\Future$ in state
$\msg\TagFULL\x$ accepts a $\TagGet$ operation carrying an argument
$\User$. The reaction leaves the state of the object unchanged (the
message $\msg\TagFULL\x$ is restored into the $\Future$'s mailbox)
and additionally stores $\msg\TagReply\x$ into $\User$'s mailbox,
from which the user of the completable future variable can retrieve
the value of $\x$.
The last line of the definition \eqref{eq:future} acts as a
constructor that initializes $\Future$ to state $\TagEMPTY$.

As defined, $\Future$ does not react to message patterns of the form
$\TagEMPTY \parop \msg\TagGet\User$ or
$\msg\TagFULL\x \parop \msg\TagPut\y$. After all, an uncompleted
future variable cannot provide its value and a completed future
variable should not be completed again. Nonetheless, the two
scenarios differ crucially: trying to read an uncompleted future
variable is \emph{alright} (simply, the request will remain pending
until the variable is completed), whereas trying to complete a
future variable twice is just \emph{wrong}.  To tell the two cases
apart we need a protocol specification for completable future
variables.
The protocols (or types) $\E$, $\F$ we consider are generated by the
grammar
\[
  \E, \F ~~::=~~ \tzero ~~\mid~~ \tone ~~\mid~~ \Tag ~~\mid~~ \tstar\Tag ~~\mid~~ \E \tor \F ~~\mid~~ \E \tand \F
\]
and are akin to regular expressions defined over tags $\Tag$, except
that iteration is limited to single tags and `$\tand$' is
shuffling. \REVISED{Formally, we define the semantics of a protocol
  as the set of strings of tags inductively generated by the
  equations below
\[
  \begin{array}{@{}r@{~}c@{~}l@{}}
    \sem\tzero & = & \emptyset
    \\
    \sem\tone & = & \set{ \varepsilon }
  \end{array}
  \qquad
  \begin{array}{@{}r@{~}c@{~}l@{}}
    \sem\Tag & = & \set\Tag
    \\
    \sem{\tstar\Tag} & = & \set{ \Tag^n \mid n \in \natset }
  \end{array}
  \qquad
  \begin{array}{@{}r@{~}c@{~}l@{}}
    \sem{\E \tor \F} & = & \sem\E \cup \sem\F
    \\
    \sem{\E \tand \F} & = & \set{ u_1v_1\cdots u_nv_n \mid u_1\cdots u_n\in\sem\E, v_1\cdots v_n \in \sem\F }
  \end{array}
\]
where $\varepsilon$ denotes, as usual, the empty string, $\Tag^n$ is
the string made of $n$ occurrences of $\Tag$, and we write $u_i$ and
$v_j$ to denote arbitrary, possibly empty strings of tags.
A type specifies the legal ways of using an object.
In particular, the only legal way of using an object of type $\Tag$
is to send an $\Tag$-tagged message to it, whereas an object of type
$\tstar\Tag$ can be sent an arbitray number of $\Tag$ messages. An
object of type $\E \tor \F$ can be used either as specified by $\E$
or as specified by $\F$, whereas an object of type $\E \tand \F$
must be used as specified by both $\E$ and $\F$. For example, an
object of type $\TagEMPTY \tand \TagPut$ must be sent both an
$\TagEMPTY$ message and also a $\TagPut$ message, in whichever
order.
The constants $\tzero$ and $\tone$ have a subtle semantics.  The
only legal way of using an object with type $\tone$ is \emph{not
  using it}, because $\sem\tone$ contains the empty string
only. Concerning $\tzero$, there is no legal way of using an object
with that type, because $\sem\tzero$ is empty.  Even \emph{not
  using} such an object is illegal! This seemingly bizarre
interpretation of $\tzero$ is actually key in the construction of
the matching automaton, as we will see in
Section~\ref{sec:automaton}.  In the following, we write $\E = \F$
if $\sem\E = \sem\F$. In particular, $\E \tand \F = \F \tand \E$.
By using \emph{shuffling} instead of sequential composition we
deprive types of any information concerning the \emph{order} in
which messages are supposed to be sent to objects. There are two
motivations for this choice. First, it is often impossible to
determine the order of messages sent to an object by concurrent
(hence, independent) processes.  Shuffling allows us to easily
overapproximate an object protocol while ignoring any ordering
constraint. Second, we will use types in the construction of the
matching automaton, whose main purpose is to detect when a reaction
can fire. According to the semantics of the Objective Join
Calculus~\cite{FournetLaneveMarangetRemy03}, only the
\emph{presence} -- not the order -- of messages in the mailbox
matters in this respect.}

\REVISED{As an example of object type,} we define $\FutureType$
thus:
\begin{equation}
  \label{eq:future.type}
  \FutureType \eqdef \tstar\TagGet \tand (\TagEMPTY\tand\TagPut \tor \TagFULL)
\end{equation}

This type specifies that a completable future variable is always
either $\TagEMPTY$ or $\TagFULL$ (\cf the `$\tor$' connective), that
it can be completed once when it is $\TagEMPTY$ (\cf the innermost
`$\tand$' connective), and that it may receive an arbitrary number
of $\TagGet$ messages (\cf the `$\tstar$' connective) regardless of
its state (\cf the outermost `$\tand$' connective, allowing any
interleaving of the $\TagGet$ messages with respect to all the other
messages).

We assume a well-formedness condition on types, requiring that every
starred tag occurring in them does not occur also unstarred. This
assumption simplifies the technical development and does not appear
to impact the expressiveness of types in describing interesting
object protocols.


%% file: automaton.tex
\section{From join patterns to matching automata}
\label{sec:automaton}

\begin{figure}
  \includegraphics[width=\textwidth]{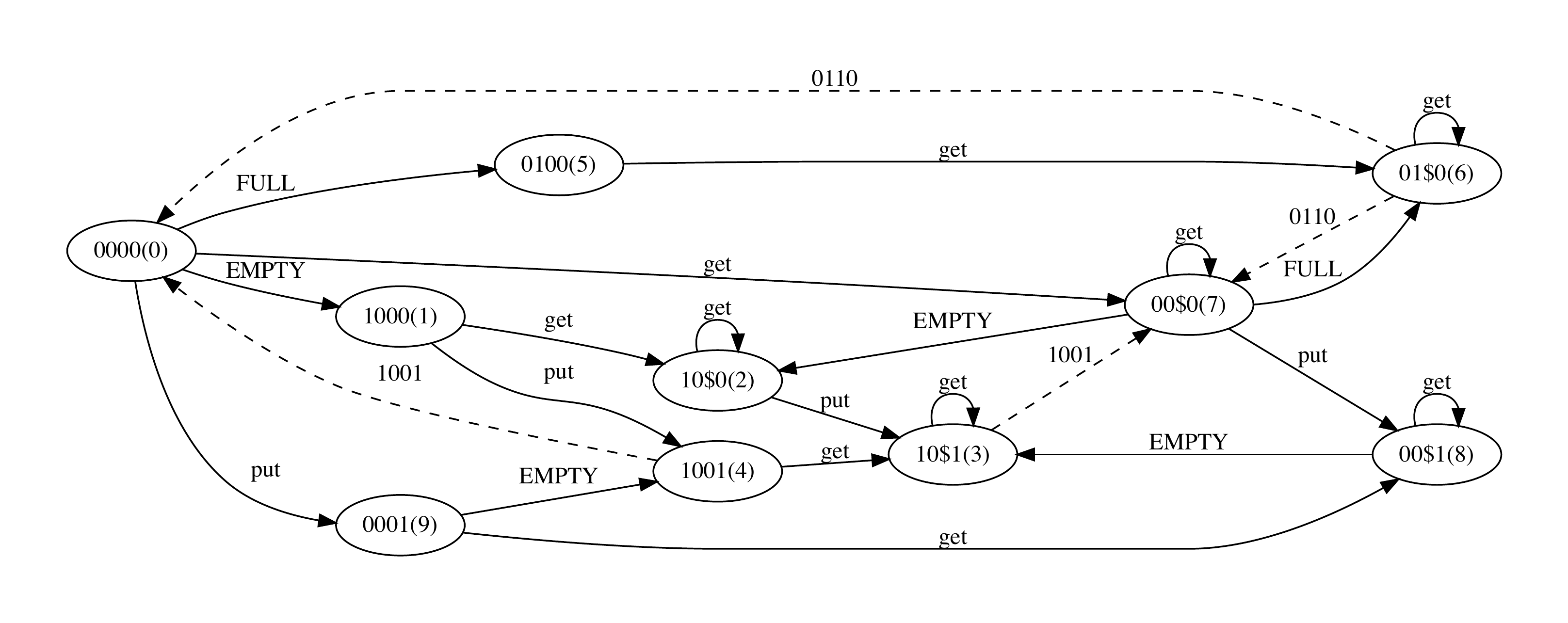}
  \caption{\label{fig:future} Matching automaton for the join
    patterns of the future variable.}
\end{figure}

In this section we describe a compilation scheme from join patterns
to \emph{matching automata}. A matching automaton is a particular
kind of finite-state automaton (FSA) whose purpose is to efficiently
detect when one of the object's reactions can fire and if a protocol
violation has occurred.  To do so, the matching automaton keeps
track of the state of the object's mailbox as messages are stored in
it.
The compilation scheme we are about to illustrate is close to the
one given by Le Fessant and Maranget~\cite{LeFessantMaranget98},
except that we use the behavioral type associated with a join
pattern to prune the state space of the resulting FSA.
We illustrate the compilation scheme using the definition of
$\Future$ \eqref{eq:future} as guiding example. The resulting
matching automaton is shown in Figure~\ref{fig:future}.

From the join pattern being compiled we determine a \emph{signature}
$\Signature$ of message tags, those understood by the object, which
is also the alphabet of the resulting matching automaton. The
signature usually coincides with the set of tags occurring in the
type of the object.  In the case of $\Future$ we have
$\Signature = \set{ \TagEMPTY, \TagFULL, \TagGet, \TagPut }$.
Tags are assumed to be \emph{totally ordered}, for instance
lexicographically.

\newcommand{\atleast}[1]{\$}

The \emph{states} of the automaton are $n$-tuples of \emph{counters}
$\counter_1\cdots\counter_n$ that provide an approximate description
of the content of the object's mailbox. The length $n$ of the tuple
is the cardinality of $\Signature$ and the $i$-th counter $\alpha_i$
of the tuple is an element of $\natset \cup \set{\atleast1}$ that
corresponds to the $i$-th tag $\Tag_i \in \Signature$ according to
the tag total order. A counter $\alpha_i \in \natset$ means that
there are \emph{exactly} $\alpha_i$ messages with tag $\Tag_i$ in
the mailbox, whereas a counter $\alpha_i = \atleast1$ means that
there is \emph{at least} one message with tag $\Tag_i$ in the
mailbox.
We use the behavioral type of the object to determine the form of
counters. We say that a message $\Tag$ is \emph{unbounded} in $\E$
if $\tstar\Tag$ occurs in $\E$ and that it is \emph{bounded}
otherwise.  It is easy to show that, for every bounded message
$\Tag$ in $\E$, there exists $N\in\natset$ such that every trace in
$\sem\E$ contains at most $N$ occurrences of $\Tag$, in which case
we say that $\Tag$ is $N$-bounded.
We choose counters of the form $k\in\set{0,\dots,N}$ for $N$-bounded
messages and of the form $0$ or $\atleast1$ for unbounded
messages. \REVISED{As we will see shortly, the distinction between
  bounded and unbounded messages is needed for handling the
  automaton's transitions.}
In $\FutureType$, $\TagEMPTY$, $\TagFULL$ and $\TagPut$ are
$1$-bounded and $\TagGet$ is unbounded.
The \emph{initial state} of the automaton is $0\cdots0$, describing
an empty mailbox.
For ease of reference, the states in Figure~\ref{fig:future} are
followed by a unique index in parentheses.

The automaton has two kinds of transitions. \emph{Receive
  transitions} (the solid arrows in Figure~\ref{fig:future})
correspond to the arrival of a message in the object's mailbox. For
example, the future's automaton has an $\TagEMPTY$-tagged transition
from state $0000$ to state $1000$ recording the fact that, if the
future's mailbox is empty and an $\TagEMPTY$ message arrives, we end
up in a mailbox configuration that has exactly one $\TagEMPTY$
message and no other message.
\emph{Consume transitions} (the dashed arrows in
Figure~\ref{fig:future}) correspond to the firing of a reaction,
which causes the consumption of the messages occurring in the join
pattern from the object's mailbox. \REVISED{Consume transitions are
  slightly more difficult to handle than receive transitions. The
  residual number of bounded messages after a consume transition can
  be computed exactly by decrementing the corresponding counters in
  the starting state. On the contrary, the counter describing the
  residual number of an unbounded message after a consume transition
  can be 0 or $\atleast1$ depending on whether all such messages
  have been consumed or if at least one remains in the mailbox.}
For example, in Figure~\ref{fig:future} there are two such
transitions from state $01\atleast10$ leading to states $0000$ and
$00\atleast10$ depending on whether the consumed $\TagGet$ message
was the last one with such tag in the mailbox or not. The
implementation of the automaton deterministically chooses which
consume transition to follow by means of a runtime
check. \REVISED{The fewer unbounded messages there are in a type,
  the fewer runtime checks are necessary, the more efficient the
  code of the matching automaton is} (more on this in
Section~\ref{sec:generator}).
We label consume transitions with an $n$-tuple indicating which
messages are consumed, hence the reaction that fires.

This basic construction admits a few refinements which help reducing
the size of the resulting FSA.
For example, Le Fessant and Maranget~\cite{LeFessantMaranget98}
observe that, under the assumption that a reaction fires \emph{as
  soon as} possible, receive transitions departing from states with
outgoing consume transitions are useless. Indeed, as the messages
consumed by the reaction are removed from the mailbox, the automaton
will move to a state describing a mailbox configuration with
\emph{strictly fewer} messages.
As we will see in Section~\ref{sec:generator}, this refinement is
not applicable in our setting because our implementation of the
matching automaton does not always guarantee such prompt firing of
transitions.
Another refinement not considered by Le Fessant and
Maranget~\cite{LeFessantMaranget98} is made possible by the use of
behavioral types. The attentive reader may have noticed that there
is no state corresponding to the tuple $1100$ in the FSA of
Figure~\ref{fig:future}. This is not an oversight: the tuple $1100$
describes a mailbox configuration containing \emph{both} an
$\TagEMPTY$ message \emph{and also} a $\TagFULL$ message, but this
configuration is not found in $\sem\FutureType$. If such
configuration is reached, the object should notify the user (\eg
through an exception) of the fact that its protocol has been
violated rather than trying to handle the situation in some way.

To identify \emph{illegal states} we need an operator on types that
is closely related to \emph{Brzozowski derivative} for plain regular
expressions~\cite{Brzozowski64} and that is inductively defined
thus:
\[
  \begin{array}{@{}r@{~}c@{~}l@{}}
    \der\tzero\Tag & = & \tzero
    \\
    \der\tone\Tag & = & \tzero
  \end{array}
  \quad
  \der{\Tag{}}{\Tag'} =
  \left\{
    \begin{array}{@{}l@{\quad}l@{}}
      \tone & \text{if $\Tag = \Tag'$}
      \\
      \tzero & \text{otherwise}
    \end{array}
  \right.
  \quad
  \der{(\tstar\Tag)}{\Tag'} =
  \left\{
    \begin{array}{@{}l@{\quad}l@{}}
      \tstar\Tag & \text{if $\Tag = \Tag'$}
      \\
      \tzero & \text{otherwise}
    \end{array}
  \right.
  \quad
  \begin{array}{@{}r@{~}c@{~}l@{}}
    \der{(\E\tor\F)}\Tag & = & \der\E\Tag \tor \der\F\Tag
    \\
    \der{(\E\tand\F)}\Tag & = & \der\E\Tag\tand\F \tor \E\tand\der\F\Tag
  \end{array}
\]

In words, $\der\E\Tag$ describes the language of traces obtained by
removing a single occurrence of $\Tag$ from those traces of $\E$
that have at least one. Note that $\der\E\Tag = \tzero$ if no trace
of $\E$ contains at least an occurrence of $\Tag$. As we will see
shortly, this property is key for identifying illegal states.

The next step is to annotate each state of the FSA with a type, as
follows: the initial state is annotated with the type assigned to
the object by the programmer. If a state $s_1$ annotated with $\E$
has an outgoing transition labeled $\Tag$ to another state $s_2$,
then $s_2$ is annotated with $\der\E\Tag$. Since there may be
distinct paths leading to the same state in the FSA, one may wonder
whether this annotation procedure is well defined. It is possible to
show that the annotation for each state is unique modulo type
equivalence. This follows from three facts: (1) the FSA is
constructed in such a way that distinct minimal paths (of receive
transitions) between two states only differ for the order of tags,
(2) the order of derivatives is irrelevant, namely
$\E[\Tag][\Tag'] = \E[\Tag'][\Tag]$ for all $\E$, $\Tag$ and $\Tag'$
and (3) type well-formedness ensures that $\der\E\Tag = \E$ if
$\Tag$ is unbounded in $\E$.
To illustrate, let us annotate a few states of
Figure~\ref{fig:future}.
The initial state is annotated with $\FutureType$.
From this state we can reach the state $1000$ with an
$\TagEMPTY$-labeled transition, hence we annotate $1000$ with
$\FutureType[\TagEMPTY] = \tstar\TagGet \tand \TagPut$. From $1000$
we can reach the state $1001$ with a $\TagPut$-labeled transition,
hence we annotate $1001$ with
$\FutureType[\TagEMPTY][\TagPut] =
(\tstar\TagGet\tand\TagPut)[\TagPut] = \tstar\TagGet$. Note that
there is another path from $0000$ to $1001$ labeled with
$\TagPut,\TagEMPTY$ and that
$\FutureType[\TagPut][\TagEMPTY] = \tstar\TagGet$.
Now consider the state $1100$. This state could be reached from
$0000$ along a path labeled with $\TagEMPTY,\TagFULL$.  We compute
$\FutureType[\TagEMPTY][\TagFULL] =
(\tstar\TagGet\tand\TagPut)[\TagFULL] = \tzero$. The fact that we
have obtained $\tzero$ means that the state $1100$ describes a
configuration of the mailbox that violates the type
$\FutureType$. In general, every state for which the type annotation
is (equivalent to) $\tzero$ is \emph{illegal} and can be eliminated.
Figure~\ref{fig:future} shows the FSA without illegal states.
%

The elimination of illegal states is not just an optimization aimed
at reducing the size of the resulting FSA, but plays a key role for
the soundness of our approach, which is based on a mixture of
compile-time and runtime checks. We have stated that the matching
automaton should recognize all and only those mailbox configurations
that are legal according to the behavioral type of the object so
that, in case or protocol violation, a suitable exception can be
thrown. With the above construction, a protocol violation is
promptly detected as a missing receive transition from the current
state of the FSA.


%% file: generator.tex
\section{Code generation}
\label{sec:generator}

\begin{figure}
\begin{JD}
  @Protocol("*get·(EMPTY·put + FULL)")
  class Future<A> {
    @State     private void EMPTY();
    @State     private void FULL(A x);
    @Operation public  A    get();
    @Operation public  void put(A x);
    @Reaction  private void when_EMPTY_put(A x) { this.FULL(x); }
    @Reaction  private A    when_FULL_get(A x) { this.FULL(x); return x; }
    public Future() { this.EMPTY(); }
  }
\end{JD}
\caption{\label{lst:code} Java implementation of a completable future variable (source code).}
\end{figure}

A Java programmer using our concurrent TSOP approach for
implementing a completable future variable writes the code shown in
Figure~\ref{lst:code}. What we see there is a syntactically-valid
Java class with a few Java annotations (in magenta) that are
specific to our approach. All the boilerplate code that stores
incoming messages, matches join patterns and watches for protocol
violations is automatically generated from this code.
Before looking at the generated code, we trace the correspondence
between the Java code in Figure~\ref{lst:code} and the \ojc
constructs in~\eqref{eq:future}.

The \JI{@Protocol} annotation on line~1 specifies the behavioral
type associated with the class and corresponds to the type
$\FutureType$ in~\eqref{eq:future}. As we have discussed in
Section~\ref{sec:automaton}, this type is necessary to build the
matching automaton corresponding to the join patterns in the
class. The type refers to the name of the Java methods declared on
lines~3--6, which correspond to messages in the \ojc.  Sending a
message to an object in the \ojc amounts to invoking the
corresponding method in Java where the arguments of the method are
those of the message, with some exceptions discussed shortly. The
body of these methods will be generated automatically.

Methods corresponding to messages have either a \JI{@State} or an
\JI{@Operation} annotation. Methods of the first kind, such as
\JI{EMPTY} and \JI{FULL}, model state messages from which we do not
expect to receive an answer. For this reason, the return type of
state methods is always \JI{void}. Methods of the second kind, such
as \JI{get} and \JI{put}, model operations on the object from which
we (usually) expect to receive an answer (possibly just a signal
meaning that the operation is complete).  In the \ojc, messages like
$\TagGet$ carry an explicit continuation that the object uses to
answer the request. The Java idiom to communicate results is through
returned values instead of explicit continuations. For this reason,
operation methods usually have one less argument compared to the
corresponding message and their return type is different from
\JI{void}, as in the case of \JI{get}.
The \JI{put} method is a notable exception. The corresponding
message does not have a continuation argument because no answer is
expected from a $\TagPut$ message. However, our code generator
relies on the fact that each join pattern combines \emph{exactly
  one} operation method and zero or more state methods. Because of
this requisite (also found in
\comega~\cite{BentonCardelliFournet04}), \JI{put} is qualified as an
operation method even if does not return any significant result. The
\JI{void} return type is the obvious choice for operation methods
like \JI{put} that do not return anything.

Methods defined on lines~7--8 have a \JI{@Reaction} annotation,
meaning that they correspond to reactions in the \ojc. The name of a
reaction method reveals the structure of the join pattern it
represents and is built from the \JI{when} prefix followed by the
tags of the messages in the join pattern separated by
underscores. So, the method names \JI{when_EMPTY_put} and
\JI{when_FULL_get} respectively correspond to the join patterns
$\TagEMPTY \parop \msg\TagPut\x$ and
$\msg\TagFULL\x \parop \msg\TagGet\User$. The argument list of a
reaction method is the concatenation of the argument lists of the
joined messages (without explicit continuations), whereas its return
type is that of the one and only operation method that appears in
the pattern. For example, \JI{when_FULL_get} has a single argument
(coming from \JI{FULL}) and return type \JI{A} (the same of
\JI{get}).
The body of a reaction method is the Java transposition of the
corresponding process in the \ojc reaction. The body of
\JI{when_EMPTY_put} changes the state of the object to \JI{FULL} by
invoking the corresponding method. The body of \JI{when_FULL_get}
restores the state of the object to \JI{FULL} and then \JI{return}s
the content of the future variable to the client.
The constructor (line~9) is unremarkable.

We use Java access modifiers to control the visibility of methods:
methods corresponding to state messages are \JI{private} to enforce
the fact that, as in the original presentation of
TSOP~\cite{AldrichEtAl09}, state transitions can only be triggered
from within the class; methods corresponding to operations are
\JI{public}, for these provide the public interface to the object;
reaction methods are \JI{private}, since they will be invoked by the
automatically generated code that fires reactions (more on them
later).

\begin{figure}[t]
  \begin{minipage}[t]{0.48\textwidth}
  \begin{JD}[name=gen,gobble=3,firstnumber=auto]
    private ReentrantLock lock; 
    private Condition try_get;
    private Condition try_put;
    private int state = 0;
    private A queue_FULL = null; 
    private int queue_get = 0; 

    private void FULL (A x) {
      lock.lock(); 
      queue_FULL = x; 
      switch (state) {
      case 0: 
        state = 5;
        lock.unlock();
        break; 
      case 7: 
        state = 6;
        try_get.signal();
        lock.unlock();
        break; 
      default: 
        lock.unlock();
        throw new IllegalStateException(); 
      }
    }
  \end{JD}
\end{minipage}
\hfill
\begin{minipage}[t]{0.48\textwidth}
  \begin{JD}[gobble=3,firstnumber=last]
    public A get () {
      lock.lock(); 
      queue_get++; 
      switch (state) {
      case 0: state = 7; break;
      case 1: state = 2; break;
      case 4: state = 3; break;
      case 5: state = 6; break; 
      case 9: state = 8; break;
      default: break; 
      } 
      while (true) 
        switch (state) {
        case 6: { 
          final A x = queue_FULL; 
          queue_FULL = null;
          queue_get--; 
          state = queue_get == 0 ? 0 : 7; 
          lock.unlock(); 
          return when_FULL_get(x); 
        }
        default:
          try_get.awaitUninterruptibly(); 
        } 
    }
  \end{JD}
\end{minipage}
\caption{\label{fig:generated.code} Java implementation of a future variable (generated code).}
\end{figure}

We now illustrate how our generator expands the source class in
Figure~\ref{lst:code} into a fully functional class. Because of
space limitations, we can only focus on a few bits of generated
code, shown in Figure~\ref{fig:generated.code}.

First of all, the generator adds some fields
(lines~\ref{lst.fields.lock}--\ref{lst.fields.queue.get}) to enforce
exclusive access to instances of the class, to represent the state
of the matching automaton, and to represent message queues.
Mutual exclusion is guaranteed by a reentrant \JI{lock} (line~1) and
by condition variables associated with the operation methods
(\JI{try_get} and \JI{try_put} on lines~2--3). It would also be
possible to use implicit locks and the built-in synchronization
facilities of Java, but using explicit locks is more flexible and
sometimes results in better performing code.
The \JI{state} of the FSA is represented as a field of type \JI{int}
that contains the indexes shown within parentheses in
Figure~\ref{fig:future}. Its initial value 0 corresponds to the
initial state of the FSA.
The concrete representation of message queues depends on whether
messages are bounded and/or have arguments.
Bounded messages without arguments do not need a queue, for all we
need to know is their number in the mailbox and this number is
encoded accurately in the state of the FSA. For this reason, there
are no fields corresponding to \JI{EMPTY} and \JI{put} messages.
For unbounded message without arguments the queue is just a counter
field of type \JI{int} that tracks the number of those messages in
the mailbox. Note that \JI{get} is one of such messages
(line~\ref{lst.fields.queue.get}). Indeed, the argument of the
\JI{get} message is an explicit continuation that disappears in
Java.
For $1$-bounded messages with a single argument the queue coincides
with the value of the argument (whether the message is present or
not is encoded in the state of the FSA). This is the case of
\JI{FULL}, whose field has the same type as its argument
(line~\ref{lst.fields.queue.FULL}).
In all the other cases we use a real queue. There is no message with
such properties in the example we are considering, hence no example
of such field.
Analogous optimized representations for message queues have been
described in the
literature~\cite{FournetLaneveMarangetRemy03,TuronRusso11}.

We now turn the attention to the generated body of method \JI{FULL},
bearing in mind that invoking this method means sending a message
$\TagFULL$ to the object. The method enters the critical section
(line~\ref{lst.FULL.lock}) and stores its argument in the
corresponding queue (line~\ref{lst.FULL.enqueue}). The following
\JI{switch} implements the transitions of the FSA by analysing and
updating the \JI{state} field. There are two receive transitions
labeled \JI{FULL} in Figure~\ref{fig:future}.
The first one (lines~\ref{lst.FULL.begin.t1}--\ref{lst.FULL.end.t1})
moves the FSA from state 0 to state 5. Since 5 is not a firing state
(it has no outgoing consume transitions), we just leave the critical
section and quit the method.
The second transition
(lines~\ref{lst.FULL.begin.t2}--\ref{lst.FULL.end.t2}) moves the FSA
to the firing state 6. If this happens, another process has
previously invoked \JI{get} and is waiting for the future variable
to be resolved. Thus, we notify the condition variable \JI{try_get}
to wake such process before leaving the critical section. Note that,
after the notification and before the awoken process is scheduled to
run, other processes may invoke a method (\eg \JI{get}) on the
object and successfully enter the criticial section. For this
reason, our compilation technique differs from the one of Le Fessant
and Maranget~\cite{LeFessantMaranget98} in that it does not
guarantee that reactions fire \emph{as soon as possible}.
To conclude the description of the \JI{FULL} method, the
\JI{default} case of the \JI{switch} deals with protocol violations
(lines~\ref{lst.FULL.begin.default}--\ref{lst.FULL.end.default}). In
this case, the \JI{FULL} method has been invoked in a state in which
this message was not expected to arrive. The violation is notified
with an exception thrown just outside the critical section
(line~\ref{lst.FULL.end.default}).

The code generated for operation methods is more complex, because it
involves not only updating the state of the FSA, but also detecting
and executing reactions that fire. Let us have a look at the
\JI{get} method, whose body is divided into two parts. The first
part (lines~\ref{lst.get.lock}--\ref{lst.get.update}) stores the
\JI{get} message (by incrementing the corresponding counter) and
updates the state of the FSA. This part is similar in structure to
the generated body of state methods like \JI{FULL}, except for two
differences.
First, the \JI{default} case (line~\ref{lst.get.nop}) corresponds to
the loops in Figure~\ref{fig:future}, which do not change the state
of the FSA. In particular, the behavioral type that we are
considering allows \JI{get} messages to arrive at any time, hence
\JI{get} messages are never a cause of protocol violation, unlike
\JI{FULL} messages.
Second, no condition variable is notified, even if the FSA ends up
in a firing state, because it is the very process executing this
code that takes care of executing the code corresponding to the
reaction (\cf line~\ref{lst.get.nofire}).

The second half of the method contains the code that detects and
executes a firing reaction
(lines~\ref{lst.get.begin}--\ref{lst.get.end}). The state of the FSA
is repeatedly checked: as long as the state is a non-firing one, the
method suspends and waits to be awoken (line~\ref{lst.get.wait});
when a firing state is detected (line~\ref{lst.get.begin.fire}), the
messages consumed by the firing reaction are removed from the
corresponding queues and their arguments (if there are any) are
stored in local variables
(lines~\ref{lst.get.remove.FULL}--\ref{lst.get.remove.get}).
The state of the FSA is then updated to account for the removal of
messages from the mailbox of the object
(line~\ref{lst.get.update.state}). Unlike previous state updates,
this one requires a runtime check: in Figure~\ref{fig:future}, there
are two outgoing consume transitions from the firing state 6,
depending on whether the residual number of \JI{get} messages in the
mailbox is 0 or not. In the first case, the FSA goes back to the
initial state. In the second case, the FSA moves to state 7.
Finally, the method leaves the critical section
(line~\ref{lst.get.unlock}) and invokes the appropriate reaction
method (line~\ref{lst.get.fire}).


%% file: conclusion.tex
\section{Concluding Remarks}
\label{sec:conclusion}

We have presented a generative approach enabling concurrent TSOP in
Java, \REVISED{thus providing a practical implementation of the
  concurrent TSOP methodology whose theoretical foundations have
  been studied in previous work~\cite{CrafaPadovani17}.}  Our
approach makes very few assumptions on the host programming language
and is therefore easily portable to other mainstream languages.
\emph{En passant}, we have described a refinement of Le Fessant and
Maranget's~\cite{LeFessantMaranget98} compilation scheme for join
patterns using behavioral types and have shown an implementation
technique of join patterns that integrates smoothly with -- and
takes advantage of -- Java without relying on libraries or language
extensions.
Specifically, we use the native sequential composition and
\JI{return} instead of explicit continuations for imposing an order
in the execution of code and for returning the result of operations.

\REVISED{Compared to the theoretical study of Crafa and
  Padovani~\cite{CrafaPadovani17}, here we have adopted a simpler
  type language where only the tag of messages -- and not the type
  of their arguments -- is reported in an object protocol. This
  simplification is made possible by the fact that our approach
  defers typestate checking at runtime, thus rendering the type of
  message arguments unimportant. Besides, the use of sequential
  composition in place of explicit continuations reduces the need of
  message arguments with complex (behavioral) types and results in a
  more natural programming style. Another difference concerns the
  semantics of types, which is given here in terms of traces instead
  of multisets~\cite{CrafaPadovani17}. Nonetheless, the two
  semantics induce the very same notion of type equivalence.}

Plaid~\cite{SunshineEtAl11A,SunshineEtAl11B} and
StMungo~\cite{KouzapasEtAl18} are notable implementations of
TSOP. Plaid supports TSOP natively, whereas StMungo relies on
external tools for analysing annotated Java code. \REVISED{The main
  advantage of Plaid and StMungo compared to our approach is that
  they are} able to provide static protocol conformance guarantees,
whereas our approach shifts most of the typestate checking at
runtime. However, neither Plaid nor StMungo support
\emph{concurrent} TSOP.

Many ideas for further developments stem from this work. For
example, the runtime information on the state of the matching
automaton could enable forms of error recovery that notoriously
clash with static analysis. The same information might also be
useful to implement a \emph{gradual type system} for concurrent
TSOP~\cite{SiekTaha07,WolffGarciaTanterAldrich11}.
Finally, we are aware that lock-based compilation of join patterns
\emph{\`a la} Le Fessant and Maranget~\cite{LeFessantMaranget98}
does not scale well to large numbers of processes. Unfortunately,
scalable implementations of join patterns~\cite{TuronRusso11} are
not based on matching automata, making it difficult to detect
protocol violations. Whether scalability and runtime protocol
checking can be reconciled remains to be established.
